\documentstyle[prl,aps,epsfig,multicol]{revtex}
\begin{document}

\def\be{\begin{equation}}
\def\ee{\end{equation}}
\def\bea{\begin{eqnarray}}
\def\eea{\end{eqnarray}}
\def\tc{$T_c$}
\def\tcl{$T_c^{1*}$}
\def\c2{$ CuO_2$}
\def\lco{$La_2CuO_2$}
\def\123{YBCO}
\def\lsco{LSCO}
\def\bi{Bi-2201}
\def\sd{$s^\dagger_{i\sigma}$}
\def\s{$s^{}_{i\sigma}$}
\def\bd{$b^\dagger_{ij}$}
\def\b{$b^{}_{ij}$}
\def\t{$\sum_\sigma s^\dagger_{i\sigma}s^{}_{i\sigma}$}
\def\vone{\vskip 1.0cm}
\def\vtwo{\vskip 2.0cm}
\def\vthree{\vskip 3.0cm}
\def\qnf{~QNF~}

\title {Quantum Number Fractionization in Cuprates:\\
U(1) RVB Gauge Theory and Senthil Fisher Constructions}

\author{G. Baskaran }

\address{Institute of Mathematical Sciences\\
C I T Campus\\
Madras 600 113, India }

\maketitle

\begin{abstract}
We critically look at Senthil Fisher constructions of quantum number 
fractionization in cuprates. The first construction, while mathematically
correct is of limited relevance to cuprates. The second one has a 
missing U(1) symmetry.  Once this aspect is repaired, it reveals itself 
as the U(1) RVB gauge theory construction of quantum number 
fractionization. An approximate local $Z_2$ symmetry arises in the spin 
sector for a reason very different from Senthil and Fisher's. The charge 
sector continues to have the local U(1) symmetry  and the metallic spin gap 
phase of the 2d cuprates may, in principle, carry (irrational) fractional 
charge $e^* \neq e$ obeying Haldane's fractional exclusion statistics 
like the 1d models. 
\end{abstract}

\begin{multicols}{2}[]

In the RVB theory of cuprates\cite{pwascience,pwabook}the 
(s)slave-particle formalism\cite{zoupwa} and the U(1) gauge theory 
development\cite{gbgauge,wiegman,gbnobel} suggested a natural 
mechanism of quantum number fractionization 
in undoped and doped cuprates. The spin charge separation\cite{KRS}
simply corresponded to the deconfinement\cite{gbgauge} 
of s-particles carrying a new U(1) charge called RVB charge\cite{gbnobel}. 
As the theory involves matter (a fermi sea of spinons
and bosonic holons) interacting with dynamically generated U(1)
gauge fields, spinon deconfinement turned out to be a natural
(the only ?) possibility\cite{gbnobel} even in the 1 and 2 
space dimensional
t-J models (in the absence of matter the U(1) gauge theory 
in 1 and 2 space dimensions have only a confined phase)
. Soon after the formulation of the  RVB gauge theory, 
a reduction of the local U(1) symmetry to a local $Z_2$ symmetry 
was found\cite{gbyulu} leading to a theory of $Z_2$ gauge fields 
coupled to 
spinon and holon matter utilizing an important identity due to 
Marston\cite{marston}.  This approach has remained unexplored. 
Several authors, Patrick Lee\cite{plee} and collaborators 
in particular 
have pursued the gauge theory approach in the last one decade.

Recently Senthil and Fisher (SF) have advocated a construction of 
quantum number fractionization (QNF) $ 2e \rightarrow e + e $ in 
two different ways\cite{SF1,SF2}. 
In the most recent approach\cite{SF2}, SF study the t-J model in the 
s-particle formalism and split a neutral d-wave spinon pair into two
and use it to construct a spinon and a `chargon'. 
We critically analyze this construction\cite{SF2} 
and find a missing local U(1) gauge invariance. This
invariance is necessary to ensure the crucial double occupancy
constraint. We have found a way of correcting this serious  
difficulty. Once this is done, SF's construction becomes the
same as the U(1) RVB gauge theory construction 
of \qnf.  In the earlier approach\cite{SF1}, a charge `$e$' excitation of 
the charge `2e' fluid in the metallic spin gap phase of the 
cuprate is invoked.  While this in principle is an allowed possibility 
in the cuprates we will argue that this is only a particular case allowed 
by the U(1) RVB gauge theory of the t-J model; a general possibility
allowed by the U(1) gauge theory is more consistent with the `charge 
gaplessness' of the metallic spin (pseudo)gap phase.  

In the process we also bring out a $Z_2$ aspect of the RVB 
gauge theory that was discussed earlier in the game
\cite{gbyulu,marston}. A fundamental
difference of our $Z_2$ gauge symmetry, which is restricted 
to the spin sector of the doped Mott insulator, from that of SF 
is also brought out. Having established the connection of SF's
construction to the old U(1) RVB mechanism, we also point out
how the local U(1) symmetry of the charge sector (in the metallic
spin gap phase) provides a more general \qnf and the
possibility
of fractional charge for holons $e^* \neq e$. In this way we
also suggest that the physics of the charge sector is fairly 
dimension independent. We also point out that in a (gapless) 
metallic 2d state such as the spin gap phase with extended 
(in general having power law form factor) spinons and holons it is 
more meaningful to talk in terms of the Haldane's exclusion 
statistics\cite{haldaneexcl} rather than limit oneself to ideas 
derived from exchange statistics.

First we elaborate how the U(1) RVB gauge theory allows for a 
unique `U(1) string' representation of the physical spinon and 
holon, allowing for a general possibility of charge
$e^*_{} \neq e$ for the holon. Then show how the recent construction 
of SF, when corrected, becomes the RVB theory's U(1) string
representation.

In the (s)slave-particle representation of the the t-J model 
an electron is decomposed as
\be
c^\dagger_{i\sigma} \equiv e_i s^{\dagger}_{i\sigma}
\ee
in terms of the `bare spinon' and `bare holon' operators 
(\sd,$e^\dagger_i$). The double occupancy constraint takes the
form $ \sum_\sigma$ \sd \s + $e^\dagger_i e^{}_i = 1 $.  
All physical objects are invariant under
the local U(1) gauge transformation applied to the s-particles:
\be
(s^\dagger_{i\sigma}, e^\dagger_i) \rightarrow
e^{-i\theta_i} (s^\dagger_{i\sigma}, e^\dagger_i)
\ee
In view of the above transformation property, we say that a 
bare spinon or a holon carry an unit of U(1) `charge' which we 
called RVB charge\cite{gbnobel}. These constituent s-particles are not 
physical particles.  Physical particles must remain invariant
under the local U(1) gauge transformation. Gauge theory 
suggests that we attach electric field lines to `electric
charges' to make them gauge invariant. In our system the field 
lines that go to infinity (boundary of the sample) can be created 
using spin singlet `U(1)string' operators such as:
\be
U_{li_1}U_{i_1 i_2} U_{i_2 i_3}...  \equiv U_l(C)
\ee
The link variables $U_{ij}$ are chosen from the three types of 
spin singlet operators:
$ U_{ij} =  b^\dagger_{ij}
 = {1\over{\sqrt 2}} (s^\dagger_{i\uparrow} s^\dagger_{j\downarrow}
- s^\dagger_{i\downarrow} s^\dagger_{j\uparrow})$,
or $\tau_{ij}  = \sum_\sigma $\sd $s^{}_{\j\sigma}$ 
or$~~ e^\dagger_ie_j$ and their complex conjugates.
A simple example of the string operator used in ref\cite{gbgauge,gbnobel}for
the Mott insulator is $ b^{}_{l i_1} b^\dagger_{i_1 i_2} 
b^{}_{i_2 i_3} . . . . .~~$ . 
By construction, under the local U(1) transformation a U(1) string 
operator transforms as $~~U_{li_1}U_{i_1 i_2} U_{i_2 i_3} ~~...~~ 
\rightarrow e^{i\theta_l} U_{li_1}U_{i_1 i_2} U_{i_2 i_3}~~...~~$.
The field line operators represent the `spin singlet'
disturbance and charge disturbance created in the spin liquid 
state in the process of \qnf,  of transporting a partner 
spin-$\frac{1}{2}$ moment to the boundary of the system
along the sequence C (not necessarily formed by nearest neighbor
lattice sites).
In a continuum approximation of the U(1) RVB gauge theory 
the string operators  correspond to: $e^{i\int_{\bf r}^\infty
{\bf A}_{rvb}\cdot d{\bf l}}$. Here ${\bf A}_{rvb}$ is the
dynamically generated U(1) RVB gauge field. That is, 
schematically a spin one operator or an electron operator undergo the 
following change during quantum number fractionization:
\bea
S_l^+ & = & s^\dagger_{l\uparrow} s^{}_{l\downarrow} \rightarrow
s^\dagger_{l\uparrow}  U_{li_1}U_{i_1 i_2} U_{i_2 i_3}~~...~~
U_{i_n\infty} s^{}_{\infty\downarrow}   \nonumber \\
c^\dagger_{l\sigma}  & = & e^{}_{l} s^\dagger_{l\sigma} \rightarrow
e^{}_{l}  U_{li_1}U_{i_1 i_2} U_{i_2 i_3}~~...~~
U_{i_n\infty} s^\dagger_{\infty\sigma}   \nonumber
\eea
The field line operators in combination with the bare 
s-particles have the full local U(1) gauge invariance and
provides {\em the most general} localized physical spinon 
or holon operator characterized by a string C and site l:
\be
\zeta^{s\dagger}_{l\sigma} = s^\dagger_{l\sigma} U_l(C) 
~~~{\mbox and}~~~
\zeta^{h\dagger}_{l} = e^\dagger_{l\sigma} U_l(C) 
\ee
A propagating low energy spinon or holon eigen state of a spin
liquid state exhibiting quantum number fractionization
(deconfinement) can be created by the operators:
\bea
\zeta_{\sigma}^{s\dagger}({\bf q}) & \equiv & \sum_{l,C} 
e^{i{\bf q}\cdot {\bf r}_l}
\phi^{s}(C,l;{\bf q}) s^\dagger_{l\sigma} U_l(C) \nonumber \\
   & \equiv  & \sum_l 
e^{i{\bf q}\cdot {\bf r}_l} s^\dagger_{l\sigma} \eta^{s}_l({\bf q}) 
~~~{\mbox and}~~~ \nonumber \\
\zeta^{h\dagger}({\bf q}) & \equiv & \sum_{l,C} 
e^{i{\bf q}\cdot {\bf r}_l}
\phi^{h}(C,l;{\bf q}) e^\dagger_{l} U_l(C) \nonumber
\\
   & \equiv  & \sum_l 
e^{i{\bf q}\cdot {\bf r}_l} e^\dagger_{l} \eta^{h}_l({\bf q}) 
\eea
The string amplitudes $\phi^{s}(C,l;{\bf q})$ and 
$\phi^{h}(C,l;{\bf q})$ depends on the form of the Hamiltonian
and in general may contain some further quantum numbers such
as `chirality' for the spinon and holon. 

The operator $\eta^{s,h}_l = \sum_C  \phi^{s,h}(C,l;{\bf q}) U_l(C)$  
creates a quantum delocalized U(1) string excitation starting 
from a site $l$; in a classical picture it is a radially flowing 
U(1) field lines emanating from a site $l$.
We call $\eta^{s}_l$ and $\eta^{h}_l$ as the `radial field operators' 
(RFO) of the deconfined phase for the spinon and holon. 
By construction the RFO's  transform as
$\eta^{}_l \rightarrow e^{i\theta_l} \eta^{}_l$
under the local U(1) gauge transformation. 

In a deconfined state the RFO represents the `radial' fashion in
which \qnf takes place leading to the possibility of a finite 
energy spinon or holon. The local U(1) gauge invariant physical 
spinon operator at a site $l$, $s^\dagger_{l\sigma}\eta^s_l$ 
and the holon operator $e^\dagger_{l}\eta^h_l$ are the 
analogue of the finite energy Laughlin quasi hole operator. 
They create a spin half moment or a missing / excess  charge 
in an otherwise singlet environment, by pushing another spin 
half moment to the boundary. 

In a gapless deconfined state the spin and charge disturbance 
created (by the RFO), in the process of \qnf,
will extend to infinity in a power law fashion. We will call this 
as a power law form factor of the `extended' spinon or holon.  
The form factors will in general be different 
for a spinon and a holon.  In the 1d Heisenberg and Hubbard 
models there is a domain wall of spin disturbance of infinite
size, around the unpaired spin, as the energy of the spinon 
tends to zero. 

Also the spin SU(2) symmetry (present for 
zero magnetic field) allows only half integer value of the
spin after the quantum number fractionization. In terms of RFO 
the virtual excitations they create are always of singlet type.  
Once we have an external magnetic field the spin rotation
symmetry is reduced to U(1) symmetry and the spin of the
spinon can be different from half, as is also known 
to be the case for the 1d Heisenberg model. In the presence of the 
magnetic field the RFO will contain some triplet components, 
which may change the effective spin of the spinon from half.

The case of holon is more complex in the presence of gapless  
charge excitation, as in the spin gap metallic phase. 
The U(1) symmetry associated with the electric charge in 
general allows any fraction $~e^*_{} \neq e~$ for the holon 
excitations. This is formally achieved by the appropriate 
choice of the `pure' charge fluctuation terms 
$e^\dagger_{i_s}e^{}_{ i_{s+1}}$ in the U(1) string, 
which can push or attract a fractional
average charge away from the site $l$. This is indeed what
happens in the 1d Hubbard model. In general, in view of the 
charge gaplessness of the metallic phase of the 2d cuprates, 
the charge of the holon should continuously changes as a function 
of doping $x$ from $e^*_{} = e$ for the Mott insulator. I do not 
find a basic principle or a topological reason in the metallic 
phase of the cuprates which restricts the charge $e^*_{}$ 
to $e$. The power law accumulation of charges and the associated
pushing of states to the boundary also leads 
to the natural possibility of Haldane's fractional exclusion 
statistics.

All our discussion was for an infinite system. If we have
a finite and large system, a spinon or a holon can not be 
created in isolation, they have to be created in pairs.
Our formal construction can be carried out without any 
difficulty in this case as well.
The above discussion does not prove the existence of deconfined
phase in some 2d models, but gives a unique construction
for objects with fractional quantum numbers in the s-particle
formalism. It should be remarked that the RVB construction
is dimension independent, suggesting the likely possibility
of a fairly universal character of the quantum number 
fractionization in the t-J model.

Having discussed the essential aspects in the U(1) gauge mechanism 
of quantum number fractionization in some detail we go to the 
construction of SF. Central to their construction\cite{SF2}, 
for the t-J model in the s-particle representation is a neutral 
d -wave spinon pair operator $d_i$  and `half' of the pair $b_i$ 
defined in terms of the nearest neighbor valence bond operator 
$b^\dagger_{ij}
 = {1\over{\sqrt 2}} (s^\dagger_{i\uparrow} s^\dagger_{j\downarrow}
- s^\dagger_{i\downarrow} s^\dagger_{j\uparrow})$, 
\be
d^\dagger_i  \sim  \sum_{j} \alpha_{ij} b^\dagger_{ij} 
~~ {\mbox and} ~~~(b^\dagger_i)^2  \sim  d^\dagger_i
\ee
And $\alpha_{ij} = 1 ~~{\mbox and}~~ -1$ for the horizontal and vertical
nearest neighbor bonds respectively and zero otherwise. 
The half of spinon pair $b_i$  is used to define
the the `spinon' and `chargon' operators $(f^{}_{i\sigma}, h^{}_i)$
through the following relations:
\be 
e_i  \equiv  b_i h_i ~~~{\mbox and}~~~
s^{}_{i\sigma}  \equiv   b_i f^{}_{i\sigma}
\ee
If $b-i$ transforms as $b_i \rightarrow e^{i\theta_i} b_i$, then
$(h_i, f_{i\sigma})$ are gauge invariant physical observables.
Let us check how $b_i$ transforms.
On applying the U(1) gauge transformation (equation 2) the
$d_i$ operator transforms as:
\be
d_i \rightarrow \sum_j \alpha_{ij} e^{i\theta_i} b_{ij} e^{i\theta_j}  
~~=~~  e^{i\theta_i} \sum_j \alpha_{ij}b_{ij} e^{i\theta_j}  
\ee
The local phase factors of the U(1) transformation do not 
completely factor out. {\em That is, the d-symmetric spinon pair,
as opposed to a valence bond pair, does not carry a definite 
U(1) charge of the local symmetry}.
Consequently, the half of spinon pair $b_i$ also does not 
carry unit U(1) charge. Thus the spinon and chargon defined 
through equation (4) are not neutral under the local U(1) gauge 
transformation, making them unqualified as physical 
excitations. One may invoke a coarse gaining argument and suggest 
that if the local U(1) gauge transformation is smooth over the scale
of the lattice parameter, $d_i$ transforms as $d_i \rightarrow
e^{2i\theta_i}$. However, a coarse graining of a local symmetry  
is not allowed, as the double occupancy constraint on individual 
sites is ensured only by the local U(1) invariance.

One may be tempted to split the bond singlet 
operator as $b^{}_{i i+ \Delta} \sim b^{}_i b^{}_{i+\Delta}$,
to extract an RFO.
This is dangerous and such a split representation for the bond
operator fails to reproduce the important commutation relation:
\be
[b^\dagger_{ij}b^{}_{ij}, b^\dagger_{jk} b^{}_{jk}]
= i{\bf S}_j\cdot( {\bf S}_i \times {\bf S}_k)
\ee

Is there a way of rectifying the U(1) gauge invariance problem ? 
The answer is obvious
once we recognize that the `half d-wave spinon pair' operator $b_i$
is invented to do the same job that the RFO $\eta^s_i$ and  $\eta^h_i$  
did in the RVB construction. The only objects that carry a U(1) charge
(apart from the s-particles themselves) are the open U(1) strings or  
suitable linear combinations. Open U(1) strings alone can neturalize the 
U(1) RVB charge of $s^\dagger_{i\sigma}$ or $e^\dagger_i$. The $Z_2$
string of SF, for the t-J model is a consequence of their erroneous 
construction, and it does not solve our U(1) problem.

It is also clear that once we extract $\eta_i$ in a gauge
invariant fashion as detailed above, there is no new 
$Z_2$ symmetry - we have the full U(1) symmetry. However, for 
a reason very different from Senthil and Fisher's, we will
show below an approximate reduction of the original U(1) 
to a $Z_2$ symmetry arising from the s-particle construction. 
This is related to the `RVB flux' or `RVB magnetic field'
that came up in a concrete fashion from the discovery of 
Affleck-marston's\cite{affleck} $\pi$-flux phase of the 
Heisenberg antiferromagnet. Later works\cite{gbchiral} 
have also introduced
a related $Z_2$ symmetry in frustrated spin systems. 
This $Z_2$ symmetry is also related to the topological order
that Kivelson\cite{kivelsondimer} and collabarators were 
trying to bring out 
through their study of quantum dimer models and the phase 
string in Sutherland's analysis\cite{sutherland}
of the short range RVB state.

To understand this in detail in our context, let us go back to the 
early work by Yu Lu, Tosatti\cite{gbyulu}
and the present author where the local 
U(1) gauge theory was reduced to a local $Z_2$ gauge theory 
using an important observation of Marston\cite{marston}.
Consider a closed string made of 
$\tau_{ij}  = \sum_\sigma $\sd $s^{}_{\j\sigma}$, 
forming an elementary pleaquette:
\be
  W_{ijkl} \equiv \tau_{ij} \tau_{jk} \tau_{kl} \tau_{li}
  \sim e^{i \oint {\bf A}_{rvb}\cdot{d\bf l}} 
\ee
Marston considered the Heisenberg model and found that this 
plaquette operator satisfies the following important identity in
the physical Hilbert space:
\be
  W_{ijkl}^3 =  W_{ijkl} 
\ee
As the  plaquette loop $W_{ijkl}$ is the analogue of the
Bohm-Aharanov phase factor of the RVB magnetic field,
Marston concluded that the important RVB fluxes enclosed by
the plaquette operator are $Z_2$ in character ( $0$ or $\pi$
flux, modulo $2\pi$) and he implemented it through a Chern-Simon 
term in the U(1) gauge theory. On the other hand the authors of 
reference\cite{gbyulu}
used Marston identity in an approximate and physical way and reduced 
the gauge theory from U(1) to $Z_2$. {\em Since Marston identity involves
only the spinon s-particles, the reduction of U(1) to $Z_2$ is 
restricted to the spin sector}. The charge sector continues to have
the U(1) symmetry, but is influenced by the condensation of 
the $\pi$ flux in the spin sector as shown in ref\cite{gbyulu}.

We discuss an important limitation in Fisher et al.'s approach
in the context of cuprates. They are strongly influenced by the 
notion of the charge incompressible quantum liquid state of an 
interacting charge 2e bose fluid in 2d, where the insulating phase 
can be thought of as a vortex pair condensed state.  This could 
exhibit \qnf with a fractional charge $e^*_{} ( = \frac{2e}{2}$). 
This is a non trivial quantum fluid, where the powerful
charge-vortex duality can be invoked for further understanding.
In view of the gap, the solitonic objects have induced charges
that are `localized' - in other words they have a finite size.
In this situation the exchange statistics makes sense. And one
can use relations such as $\Phi_c \Phi_v = 1$ (analogue of 
Dirac's quantization condition derived using spatial exchanges) 
without any problem and this 
restricts allowed fractional charges to rational values
(because the vorticity is always an integer).

Experiments in cuprates show a metallic spin gap state crossing
over to a (spin)gapless metallic non fermi liquid state as we 
increase doping.
Hence a physically meaningful 
{\bf zero temperature
reference state} is a metallic state rather 
than an insulating state. 
This reference non fermi liquid state
undergoes a superconducting transition at a \tc. We have to 
discuss the quantum number fractionization issue in this reference
metallic state rather than a charge insulator at finite doping.

The `extended' character of holon with a power law 
form factor (discussed earlier) makes the exchange statistics 
lose its precise meaning and also the corresponding 
relation $\Phi_c\Phi_v
=1$.  This was shown by Haldane and Wu\cite{haldanewu} 
in the context of a suggestion of Chao et al.\cite{chao}
for fractional exchange statistics for quantized vortices in
superfluid ${}^4$He film. Even though exchange 
statistics looses its meaning, a new possibility arises, 
namely Haldane's exclusion statistics\cite{haldaneexcl}. 
This notion is meaningful in the gapless system and is 
dimension independent. In some 
approximate studies\cite{basco} it has been proposed that 
the spinons 
and holons have non trivial exclusion statistics in the 2d
t-J model.

In view of the above discussion we conclude that the \qnf 
of $e^*_{} = \frac{2e}{2}$ is an artificial restriction in the 
spin gap phase of the cuprates. The $Z_2$ character of the spin
sector leads to the interesting possibility of $\pi$ flux 
condensation
(i.e., a cross over to the spin gap phase) from the uniform 
RVB state\cite{bza}.  It is very likely that the charge 
of the holon (and its exclusion statistics) will continuously vary 
from $e^*_{} = e$ as a function of doping. This is also a fundamental 
difference of cuprates from the well gapped quantum Hall system or 
other charge insulators.

In conclusion, if quantum number fractionization occurs in cuprates 
leading to the experimentally seen non-fermi liquid 
liquid phase and the d-wave superconducting state, 
the only viable mechanism seems to be through the RVB mechanism of spin 
charge separation, which comes out in a natural fashion in
the RVB U(1) gauge theory. The variational/RVB mean field theory
method combined with Gutzwiller projection\cite{pwascience,bza}, 
or the tomographic Luttinger liquid approach\cite{pwabook} 
are different ways of solving the same complex problem.

I wish to thank R. Shankar (Madras) for discussions.

\end{multicols}


\begin{references}

\bibitem{pwascience} P.W. Anderson, Science {\bf 235} 1196 (87)

\bibitem{pwabook} P.W. Anderson, The theory of high temperature
superconductivity (Princeton University Press, NY, 1996)

\bibitem{zoupwa}Z. Zou and P.W. Anderson, Phys. Rev. 
{\bf B 37} 627 (88) 

\bibitem{gbgauge} G. Baskaran and P.W. Anderson, Phys. Rev. {\bf B37}
 580 (88) 

\bibitem{wiegman}P.B. Wiegman, Physica Scripta {\bf T27} 160 (89)
  
\bibitem{gbnobel} G. Baskaran, Physica Scripta, {\bf T27} 
53 (89); G. Baskaran and R. Shankar, J.Mod. Phys. 
Lett.{\bf B2}, 1211 (1988)      

\bibitem{KRS} S. Kivelson, J. Sethna and D. Rokshar, Phys. Rev.
{\bf B 38} 8865 (87) 

\bibitem{gbyulu} G. Baskaran, Yu Lu and E. Tosatti,
Int. J. Mod. Phys. {\bf B 1} 555 (1988).

\bibitem{marston} J.B. Marston, Phys. Rev. Lett.,
{\bf 61} 1914 (1988)

\bibitem{plee} P.B. Wiegman, Phys. Rev. Lett. {\bf 60} 821 (88); 
Z. G. Wen and A. Zee, Phys. Rev. Lett. {\bf 62} 2873 (89); 
Z. G. Wen, F. Wilczek and A. Zee, Phys. Rev. {\bf B39} 11413 (89); 
L. Ioffe and V. I. Larkin, Phys. Rev.  {\bf B 38} 8988 (89); 
P.A. Lee and N. Nagaosa, Phys. Rev. {\bf 45} 966 (92); 
X.-G. Wen and P.A. Lee, Phys. Rev. Lett.  {\bf 80} 2193 (98); 
P.A. Lee et al.  Phys. Rev. {\bf 57} 6003 (98); Dung-Hai Lee,
Phys. Rev. Lett. {\bf84}2494(00)

\bibitem{SF1} T. Senthil, M.P.A. Fisher, cond-mat/9910224;
L. Balents, M.P.A. Fisher and C. Nayak, Phys. Rev. {\bf B 60}
1654 (99)

\bibitem{SF2} T. Senthil, M.P.A. Fisher, cond-mat/0006500

\bibitem{haldaneexcl} F.D.M. Haldane, Phys. Rev. Lett.
{\bf 67} 937 (91)

\bibitem{affleck} I. Affleck and J. B. Marston, Phys. Rev.
{\bf B 37} 3774 (88) 

\bibitem{gbchiral} G. Baskaran, Phys. Rev. Lett.,
{\bf 63} 2524 (1989); N. Read and S. Sachdev, Phys. Rev.
Lett., {\bf 66} 1773 (91); S. Sachdev and N. Read,
Int. J. Mod. Phys. {\bf B5} 219 (91); X.G. Wen,
Phys. Rev., {\bf B 44} 2664 (91)

\bibitem{kivelsondimer} D. S. Rokhsar and S. A. Kivelson,
Phys. Rev. Lett. {\bf 61} 2376 (88)

\bibitem{sutherland} B. Sutherland, Phys. Rev. {\bf 38}
7192 (88)

\bibitem{haldanewu} F.D.M. Haldane and Yong-Shi Wu,
Phys. Rev. Lett. {\bf 55} 2887 (85)

\bibitem{chao} R. Y. Chao et al., Phys. Rev. Lett.,
{\bf 54} 1339 (85)

\bibitem{basco} F. Basco et al. J. Phys. Soc. Jap.,
{\bf 65} 687 (1996)     

\bibitem{bza} G. Baskaran, Z. Zou and P.W. Anderson,
Sol. St. Commn, {\bf 63} 873 (87) 
     
\end{references}
\end{document}